\documentclass[twocolumn,aps,floatfix]{revtex4}
\usepackage{amssymb}
\usepackage{graphicx}

\begin{document}
\title{Bright solitons in Bose-Fermi mixtures}

\author{Tomasz Karpiuk,$\,^1$ Miros{\l}aw Brewczyk,$\,^1$
        and Kazimierz Rz{\c a}\.zewski$\,^2$}

\affiliation{\mbox{$^1$ Instytut Fizyki Teoretycznej, Uniwersytet w Bia{\l}ymstoku, 
                        ulica Lipowa 41, 15-424 Bia{\l}ystok, Poland}  \\
\mbox{$^2$ Centrum Fizyki Teoretycznej PAN, Aleja Lotnik\'ow 32/46, 02-668 Warsaw,
           Poland}  }

\date{\today}

\begin{abstract}
We consider the formation of bright solitons in a mixture of Bose and Fermi 
degenerate gases confined in a three-dimensional elongated harmonic trap. The 
Bose and Fermi atoms are assumed to effectively attract each other whereas 
bosonic atoms repel each other. Strong enough attraction between bosonic 
and fermionic components can change the character of the interaction within the 
bosonic cloud from repulsive to attractive making thus possible the generation of 
bright solitons in the mixture. On the other hand, such structures might be in 
danger due to the collapse phenomenon existing in attractive gases. We show, 
however, that under some conditions (defined by the strength of the Bose-Fermi
components attraction) the structures which neither spread nor collapse can be 
generated. For elongated enough traps the formation of solitons is possible even 
at the ``natural'' value of the mutual Bose-Fermi ($^{87}$Rb\,-$^{40}$K in our case) 
scattering length. 

\end{abstract}

\maketitle

Quantum degenerate atomic gases offer unique opportunities for investigating nonlinear 
properties of matter waves. Atomic Bose-Einstein condensation, first observed in $1995$
\cite{BEC}, makes possible sources of sufficiently intense atomic matter waves
allowing for the observation of nonlinear effects similar to those present in
nonlinear optics. An important difference is that while the light must propagate 
through the nonlinear medium to produce the nonlinear effects, the nonlinearity is
automatically present for atomic matter waves due to interactions between atoms.
First confirmation of nonlinear atomic matter waves properties came in $1999$ when
the four-wave-mixing phenomenon was observed in the Bose-Einstein condensate of
sodium atoms \cite{4WM}. The second was a series of experiments discovering the
solitonic behavior of atomic condensates. First, it was demonstrated with the help
of the phase imprinting technique \cite{dark} that the condensate wave function for 
repulsive interactions exhibits the dark solitonic solutions \cite{Zakharov}. Next,
also bright matter-wave solitons were created experimentally but this was achieved 
by other methods. In one of them, the interactions between atoms are changed from
repulsive to attractive using the magnetically tuned Feshbach resonances technique
\cite{bright}, the other one \cite{lattice} makes use of possibility of changing the 
sign of effective mass in the periodic potential (i.e., in the optical lattice),
a concept developed in Ref. \cite{Lenz}.

In Ref. \cite{TomekPRL} we proposed a new way for generating bright solitons in
repulsive Bose-Einstein condensates. The idea was to immerse the condensate in
a degenerate gas of fermionic atoms which attract bosons. It turns out that for 
strong enough attraction between bosons and fermions the system starts to behave 
as a mixture of two effectively attractive gases even though none of the components
alone is an attractive system. Therefore the formation of bright solitons becomes
possible. These solitons are two-component peaks with much smaller number of fermions
than bosons. However, in a three-dimensional space the atomic gases with attractive
forces might show an instability due to the collapse. This is the main goal of this
paper to answer the question whether solitons in a Bose-Fermi mixture can be
generated despite of the existing instability driven by the attractive forces.

The system, we consider, is a Bose-Fermi mixture confined in a three-dimensional
trap at zero temperature. We focus on the mixture of $^{87}$Rb and $^{40}$K
atoms in their doubly spin polarized states. At zero temperature this system 
is described by the many-body wave function 
$\Psi ({\bf x}_1,...,{\bf x}_{N_B};{\bf y}_1,...,{\bf y}_{N_F};t)$,
where $N_B$ ($N_F$) is the number of bosons (fermions). To find the time
evolution of the system we impose an assumption according to which the
many-body wave function is given in terms of single-particle orbitals in
the following way
\begin{eqnarray}
&&\Psi ({\bf x}_1,...,{\bf x}_{N_B};{\bf y}_1,...,{\bf y}_{N_F};t) =
\prod_{i=1}^{N_B} \varphi^{(B)}({\bf x}_i,t)
\nonumber \\
&&\times \frac{1}{\sqrt{N_F!}} \left |
\begin{array}{lllll}
\varphi_1^{(F)}({\bf y}_1,t) & . & . & . & \varphi_1^{(F)}({\bf y}_{N_F},t) \\
\phantom{aaa}. &  &  &  & \phantom{aaaaa}. \\
\phantom{aaa}. &  &  &  & \phantom{aaaaa}. \\
\phantom{aaa}. &  &  &  & \phantom{aaaaa}. \\
\varphi_{N_F}^{(F)}({\bf y}_1,t) & . & . & . & \varphi_{N_F}^{(F)}({\bf y}_{N_F},t)
\end{array}
\right |    
\label{wavefunction}
\end{eqnarray}
In fact, it is the simplest form that preserves the appropriate symmetry conditions
for both the bosonic part and the spin-polarized fermionic subsystem.

To find the equations that govern the evolution of the single-particle wave
functions we follow the formalism based on the Lagrangian density. 
Since the fermionic sample is spin-polarized the only interactions included
are the contact boson-boson (assumed to be effectively repulsive) and 
boson-fermion (assumed to be attractive) ones. The time--dependent many--body 
Schr\"odinger equation is fully equivalent to the following Lagrangian density
\begin{eqnarray}
&&\mathcal{L} = \frac{i\hbar}{2} \Psi^* \frac{\partial \Psi}{\partial t}
- \frac{i\hbar}{2} \Psi \frac{\partial \Psi^*}{\partial t}
- \frac{\hbar^2}{2 m_B} \sum_{i=1}^{N_B} \nabla_{{\bf x}_i} \Psi^* \;
\nabla_{{\bf x}_i}  \Psi
\nonumber  \\
&&- \frac{\hbar^2}{2 m_F} \sum_{j=1}^{N_F} \nabla_{{\bf y}_j} \Psi^* \;
\nabla_{{\bf y}_j}  \Psi  - \sum_{i=1}^{N_B} V_{trap}^{(B)}({\bf x}_i) \; 
\Psi^* \Psi   \nonumber  \\
&&- \sum_{j=1}^{N_F} V_{trap}^{(F)}({\bf y}_j) \; \Psi^* \Psi
- \sum_{i<i'} V_{int}^{(BB)}({\bf x}_i-{\bf x}_{i'}) \; \Psi^* \Psi
\nonumber  \\
&&- \sum_{i,j} V_{int}^{(BF)}({\bf x}_i-{\bf y}_{j}) \; \Psi^* \Psi
\label{Lagden}
\end{eqnarray}
treated as a function of the many-body wave function
$\Psi ({\bf x}_1,...,{\bf x}_{N_B};{\bf y}_1,...,{\bf y}_{N_F};t)$
and its time and spatial derivatives.

Therefore, the many-body wave function of the form (\ref{wavefunction}) is 
inserted into the Lagrangian density (\ref{Lagden}). Next, we apply the
mean-field approximation by integrating each term of (\ref{Lagden}) over 
all but one particular bosonic or fermionic coordinate (when the interaction 
term is considered both the one bosonic and one fermionic coordinates have to 
be left out). Utilizing the orthogonality properties of the fermionic orbitals 
(as well as neglecting $1$ in comparison with the number of bosons $N_B$) one 
gets the effective Lagrangian density
\begin{eqnarray}
&&\mathcal{L'} = \frac{i\hbar}{2} N_B \varphi^{(B) *}
\frac{\partial \varphi^{(B)}}{\partial t} -
\frac{i\hbar}{2} N_B \varphi^{(B)}
\frac{\partial \varphi^{(B) *}}{\partial t}  \nonumber  \\
&&+ \frac{i\hbar}{2} \sum_{j=1}^{N_F} \varphi_j^{(F) *}
\frac{\partial \varphi_j^{(F)}}{\partial t} -
\frac{i\hbar}{2} \sum_{j=1}^{N_F} \varphi_j^{(F)}
 \frac{\partial \varphi_j^{(F) *}}{\partial t}   \nonumber  \\
&&- \frac{\hbar^2}{2 m_B} N_B \nabla \varphi^{(B) *}
\nabla \varphi^{(B)}
- \frac{\hbar^2}{2 m_F} \sum_{j=1}^{N_F} \nabla \varphi_j^{(F) *}
\nabla \varphi_j^{(F)}   \nonumber  \\
&&- N_B V_{trap}^{(B)} | \varphi^{(B)} |^2
- V_{trap}^{(F)} \sum_{j=1}^{N_F}  | \varphi_j^{(F)} |^2  \nonumber  \\
&&- \frac{1}{2} g_{B} N_B^2 | \varphi^{(B)} |^4
-g_{BF} N_B \sum_{j=1}^{N_F} | \varphi_j^{(F)} |^2 \;| \varphi^{(B)} |^2
\label{effective}
\end{eqnarray}
which depends on single-particle bosonic and fermionic orbitals and their spatial 
and time derivatives. Here, we already introduced the interaction strengths
$g_B$ and $g_{BF}$ that determine relevant interatomic interactions and are
related to the scattering lengths $a_B$ and $a_{BF}$ via formulas 
$4\pi \hbar^2 a_B /m_B$ and $2\pi \hbar^2 a_{BF} /\mu$, respectively
($m_B$ is a mass of $^{87}$Rb atom and $\mu$ is a reduced mass of $^{87}$Rb and 
$^{40}$K atoms).

Having the effective single-particle Lagrangian density one can easily derive 
the corresponding Euler-Lagrange equations  (here, $j = 1,2,...,N_F$)
\begin{eqnarray}
&&i\hbar\,\frac{\partial\varphi^{(B)}}{\partial t} = -\frac{\hbar^2}{2 m_B}
\nabla^2 \varphi^{(B)} + V_{trap}^{(B)} \, \varphi^{(B)}  \nonumber  \\
&&+\; g_B\, N_B\, |\varphi^{(B)}|^2 \, \varphi^{(B)}
+ g_{BF} \sum_{j=1}^{N_F} |\varphi_j^{(F)}|^2 \, \varphi^{(B)}
\nonumber  \\
&&i\hbar\,\frac{\partial\varphi_j^{(F)}}{\partial t} = -\frac{\hbar^2}{2 m_F}
\nabla^2 \varphi_j^{(F)} + V_{trap}^{(F)} \, \varphi_j^{(F)}
\nonumber  \\
&&+\; g_{BF}\, N_B\, |\varphi^{(B)}|^2 \, \varphi_j^{(F)}   \;.
\label{equ}
\end{eqnarray}
Neglecting the interaction between bosons and fermions results in a separation 
of the above equations; the first of them becomes then the Gross-Pitaevskii 
equation for a degenerate Bose gas whereas the rest describes the system
of noninteracting fermions. However, when the Bose and Fermi components
attract each other strongly enough the last term in the first equation can
dominate over the term describing the bosonic mean-field energy causing
effective attraction between bosons. We are going to investigate further this 
possibility.

We solve numerically the Eqs. (\ref{equ}) for a mixture of $^{87}$Rb\,-$^{40}$K
atoms confined in a three-dimensional elongated harmonic trap as well as in an
atomic waveguide (obtained by releasing the axial confinement of the harmonic trap). 
We start from getting the ground state of a three-dimensional mixture in a harmonic 
trap. To this end, we pick up the atomic orbitals at zero time as they were the
stationary solutions of Eqs. (\ref{equ}) assuming no interaction between bosons and
fermions ($a_{BF}=0$). Therefore, the bosonic wave function is the solution of
the Gross-Pitaevskii equation, whereas the fermionic orbitals are successive (with
respect to the energy) eigenstates of the harmonic potential. Because of numerical 
limitations, the number of fermionic atoms is small (at most $15$ in our case) and
since the trap is elongated only states with higher axial quantum number are populated
(the radial quantum number equals zero). Next, we adiabatically increase the 
attraction between Bose and Fermi components.

\begin{figure}[thb]
\resizebox{2.9in}{2.3in} {\includegraphics{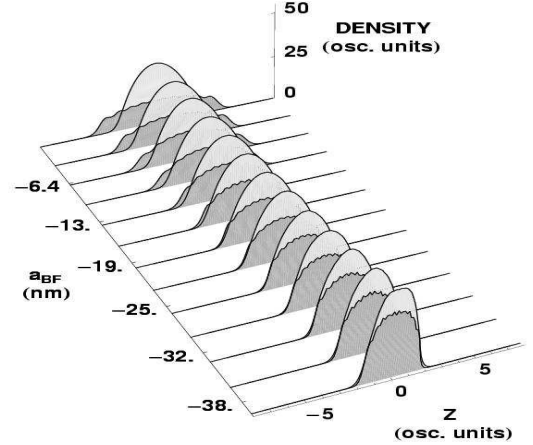}}
\caption{Axial density profiles of a three-dimensional Bose-Fermi mixture
consisting of $100$ bosons ($^{87}$Rb atoms) and $10$ fermions ($^{40}$K atoms)
confined in the harmonic trap with axial frequency $2\pi \times 1$ Hz and the
radial one $2\pi \times 300$ Hz for bosonic component (corresponding trap 
frequencies for fermions are multiplied by a factor $1.473$). Interactions 
between bosons are determined by the ``natural'' scattering length $a_B=5.8$\,nm 
whereas the strength of $^{87}$Rb atoms\,-$^{40}$K atoms attraction is being changed. 
Dark (bright) shaded areas correspond to fermions (bosons). At zero attraction 
between fermions and bosons, fermions form a broad density distribution due to the 
Pauli exclusion principle. For strong enough attraction, however, fermions are
pulled inside the bosonic cloud. } 
\label{gs10}
\end{figure}

\begin{figure}[htb]
\resizebox{2.9in}{4.in} {\includegraphics{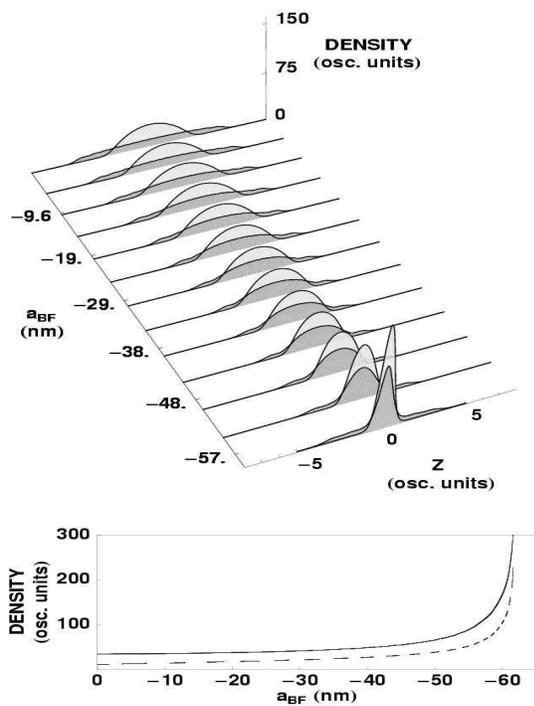}}
\caption{Illustration of a collapse phenomenon. All parameters but the number
of fermions (that equals $15$ now) are the same as in Fig. \ref{gs10}. The
lower panel shows the bosonic (solid line) and fermionic (dashed line) densities
at the center of the trap as a function of the scattering length $a_{BF}$. The
onset of the collapse at about $a_{BF}=-60$\,nm is clearly visible. } 
\label{gs15}
\end{figure}

In Fig. \ref{gs10} we plot the axial density (in units if $1/a_{ho}^3$, where
$a_{ho}=(\hbar / m_F \omega_z^{(F)})^{1/2}$) of the $^{87}$Rb\,-$^{40}$K mixture
with $100$ bosons and $10$ fermions confined in the harmonic trap. The densities
are normalized to one. The trap frequencies for bosons are $2\pi \times 1$ Hz 
($2\pi \times 300$ Hz) in axial (radial) direction (corresponding frequencies for 
fermions are multiplied by a factor $1.473$). For weaker attraction between 
bosonic and fermionic atoms fermionic cloud broadens due to the Pauli exclusion 
principle. Stronger attraction, however, can overcome the repulsion caused by the 
Pauli exclusion principle and all fermions are pulled inside the bosonic cloud. 
Increasing further the attraction results in an extremely fast growth of the density 
at the center of the trap. In other words, the collapse of the mixture happens. 
It is shown in Fig. \ref{gs15}. This phenomenon was already observed experimentally 
\cite{collapse} as well as extensively studied theoretically \cite{collthe}. It is 
well known that such a behavior is attributed to the three-dimensional realm and does 
not happen in a one-dimensional geometry.

\begin{figure}[htb]
\resizebox{2.9in}{2.3in} {\includegraphics{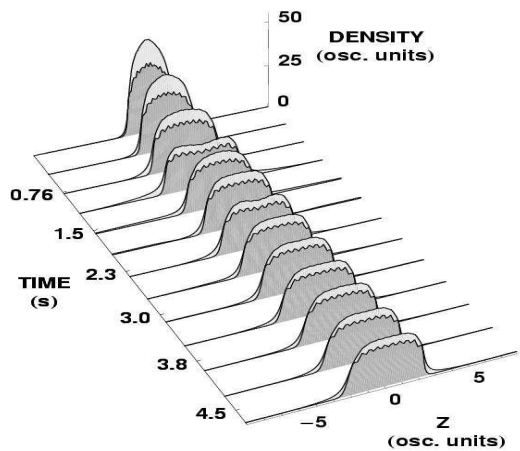}}
\caption{Two-component soliton standing at the center of the trap. The mutual
scattering length equals $a_{BF}=-40$\,nm. The axial confinement (the radial one 
is still kept on) is removed linearly within $0.5$\,s. Bosonic and fermionic 
densities fit to each other in approximately $1$s (the system looses about $6$ 
bosons within this time) and later on the single peak two-component soliton is 
formed. Radial trap frequency is the same as that in the case of Fig. \ref{gs10}.} 
\label{notrap}
\end{figure}

\begin{figure}[htb]
\resizebox{2.9in}{2.3in} {\includegraphics{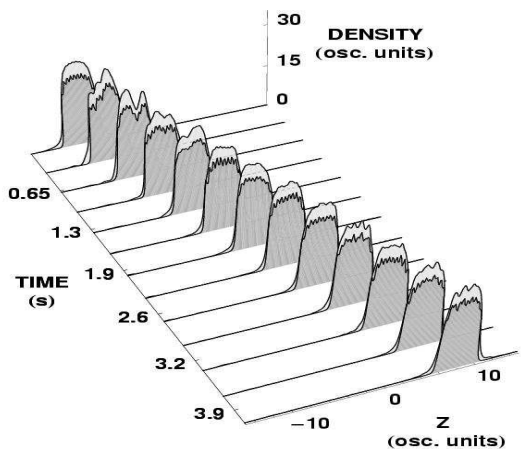}}
\caption{Illustration of the solitonic character of a structure obtained by
releasing the axial confinement as in the case of Fig. \ref{notrap}. The
structure is pushed along the axial direction (direction of no confinement) and
is moving maintaining its shape (except small ``breathing''). This is the case 
when the scattering length $a_{BF}$ lies within the window supporting the 
existence of bright solitons.} 
\label{move}
\end{figure}

\begin{figure}[htb]
\resizebox{2.9in}{2.3in} {\includegraphics{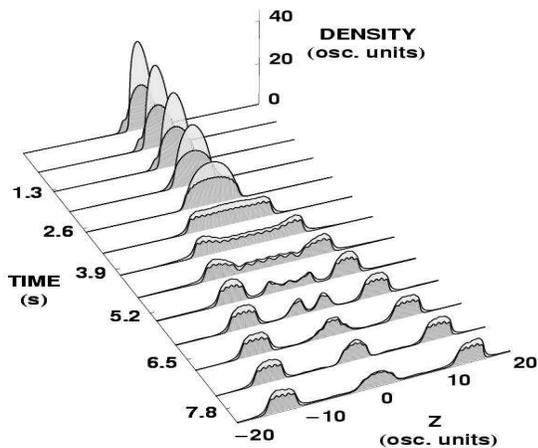}}
\caption{Nonsolitonic behavior of bosonic and fermionic clouds. The attraction
between bosons and fermions is not strong enough ($a_{BF}=-30$\,nm) and both densities
are spreading. Later on, however, the clouds break into several pieces that move 
further without changing their shapes illustrating the rule that at weaker attraction
solitons with smaller number of atoms are allowed. Here, the initial number of
fermions equals $15$.} 
\label{spreading}
\end{figure}

\begin{figure}[htb]
\resizebox{2.9in}{2.3in} {\includegraphics{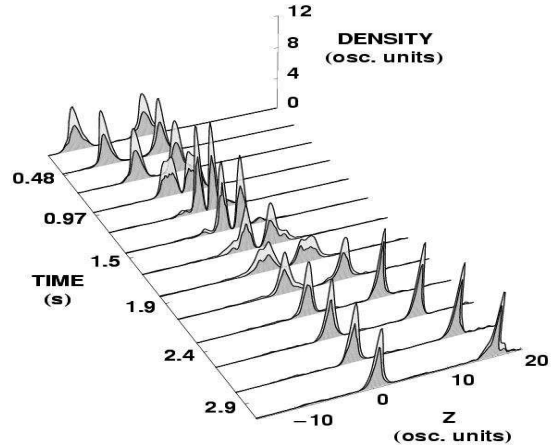}}
\caption{Collision of two solitons, each of them consisting of $6$ fermions and
$100$ bosons. The left soliton is moving towards the right one being initially
at rest. After collision (accompanied by the transfer of some atoms) the right 
soliton is moving and the left one is staying almost at rest. In this case 
$a_{BF}=-120$\,nm and the trap parameters are $2\pi \times 1$ Hz in the axial 
direction and the $2\pi \times 30$ Hz for the radial one for bosonic component.} 
\label{collision}
\end{figure}

Having the ground state of the $^{87}$Rb\,-$^{40}$K mixture in the harmonic
trap we can now reload the system to the atomic waveguide just by opening the
trap in axial direction. Results of this procedure are presented in Figs. 
\ref{notrap} and \ref{spreading}. The response of the system to the releasing
of the axial confinement depends on the value of the scattering length
$a_{BF}$. It turns out that for a particular trap geometry there is an 
interval of values of $a_{BF}$ with such a property that if $a_{BF}$ lies
within this interval the two-component single-peak structure is formed
which does not spread nor collapse. This case is shown in Fig. \ref{notrap}
where the axial confinement is linearly open in approximately $0.5$\,s. The
mixture looses $6$ bosons while both densities fit to each other and then
apparently stays in the center of the trap without changing its shape. To 
check whether the mixture forms a soliton (or at least a solitary wave) we
forced it to move by pushing along the waveguide direction the bosonic
component. It is clear from Fig. \ref{move} that bosons are able to pull
fermions and both components stay together during the movement. Thus
the mixture behaves as a solitary wave although some breathing of the structure
is observed. The axial shapes of the densities resemble rather step-like function
than the familiar solitonic secans-hyperbolicus shape. We find this shape as a
typical (i.e., present in all geometries investigated) while changing the
scattering length $a_{BF}$ by approaching the edge of the window supporting the 
existence of solitons, corresponding to the weaker attraction. For stronger
attraction the shape is ``more'' solitonic (see Fig. \ref{collision}). The
soliton in Fig. \ref{notrap} is $40$\,$\mu$m long and the bosonic density is
of the order of $10^{12}$\,cm$^{-3}$.

When the value of the scattering length $a_{BF}$ lies outside the window just
discussed (i.e., the attraction between bosons and fermions is weak enough) we
see the spreading of bosonic and fermionic clouds (Fig. \ref{spreading}). Here,
$a_{BF}=-30$\,nm and the attraction is too weak to glue the bosonic and fermionic
clouds. Both clouds spread and eventually break into several smaller droplets.
Fig. \ref{spreading} shows that at least outermost droplets (consisting of $5$
fermions and $30$ bosons) demonstrate the solitonic behavior moving without 
changing their shapes. This observation supports the view that weaker attraction 
allows only for two-component solitons with small number of atoms (as opposed to 
what is reported in Ref. \cite{Adhikari}).

Finally, we consider the collision of two bright two-component solitons within the 
waveguide (Fig. \ref{collision}). Both of them contains $6$ fermions and $100$
bosons. The right one stays at rest whereas the left one is forced to move by
pushing the bosonic component. The left soliton is moving with the velocity
$0.12$\,mm/s. Moreover, the phase $\pi$ is added to the bosonic part of the
soliton at rest, otherwise solitons merge when they meet at the center of the
waveguide and collapse. This kind of behavior is similar to what happens for 
pure bosonic solitons governed by the Gross-Pitaevskii equation. In that case
solitons repel or attract each other depending on the initial relative phase
\cite{GPsol}. If this relative phase equals $\pi$ the pure bosonic solitons always
repel each other. This is what we observe in our numerical simulations, an example
is given in Fig. \ref{collision}. After collision the right soliton starts to
move whereas the left one remains almost at rest. Usually, the transfer of some
mass happens during the collision.

To summarize our numerical work we present in Table \ref{table1} some numbers 
characterizing the formation of single-peak two-component bright solitons.
In particular, two last columns give approximate values of the critical mutual 
scattering lengths $a_{BF}^{cr}$ (the mixture spreads for the weaker attraction than 
$a_{BF}^{cr}$) and the values of $a_{BF}^{col}$ determining the onset of the collapse 
for various radial confinements. Thus, there exist windows supporting the existence of 
stable structures that maintain their shapes while moving and may survive the collisions 
(although some atoms are usually transfered during the collision). Such structures can 
be called solitons. The critical value of the scattering length can be also estimated as
in the Ref. \cite{TomekPRL}. Based on the first equation of a set of Eqs. (\ref{equ}) 
the condition for the system to enter the phase when the bosonic component becomes 
the gas of effectively attractive atoms (and hence allowing for the generation of bright 
solitons) is written as $g_B n_B=|g_{BF}^{cr}| n_F$, where $n_B=N_B |\varphi^{(B)}|^2$ 
and $n_F=\sum_i^{N_F} |\varphi_i^{(F)}|^2$ are the bosonic and fermionic densities at the
center. This criterion gives the values: $-86$\,nm, $-40$\,nm, and $-14$\,nm for the radial 
confinements: $2\pi \times 30$ Hz, $2\pi \times 300$ Hz, and $2\pi \times 1000$ Hz, 
respectively, what remains in a good agreement with the values obtained directly from the 
numerical simulations.

\begin{table}[htb]
\caption{Parameters characterizing the formation of single-peak two-component 
bright solitons.}
\label{table1}
\begin{ruledtabular}
\begin{tabular}{ccccc}
$\nu^{(B)}_r$[Hz] & $N_B$ & $N_F$ & $a_{BF}^{cr}$[nm] & $a_{BF}^{col}$[nm] \\
\hline
\hspace{0.1in}30 & 100 & 6 & -90 & -135\\
\hspace{0.1in}300  & 94 & 10 & -35 & -62\\
\hspace{0.1in}1000  & 64 & 15 & -17 & -44\\
\end{tabular}
\end{ruledtabular}
\end{table}

In Table \ref{table1} we put also the numbers of fermionic and bosonic atoms 
within the single-peak solitons. They are obtained in the following way.
For each geometry (i.e., the radial confinement) we start from the three-dimensional
harmonic trap ground state consisting of $100$ bosons and $10$ or $15$
fermions. After releasing the axial confinement the excess of atoms (bosons or
fermions) flows out of the system. For example, for the radial frequency
$\nu_r^{(B)}=30$\,Hz the mixture looses $4$ fermions and then behaves as a soliton
(to get more fermions within the peaks the initial number of bosons would have to
be bigger). For stronger radial confinement the mixture expels the excessive
bosons. We have checked that for $\nu_r^{(B)}=2000$\,Hz there exists a soliton at the 
natural value of the scattering length $a_{BF}$ (equal to $-13.2$\,nm, \cite{Jin}).
Our simulations show that at small values of $|a_{BF}|$ only solitons with small
numbers of atoms exist, the excessive bosons or fermions just leak out of the
system. Our preliminary results based on the three-dimensional variational analysis 
within the hydrodynamic model of a Bose-Fermi mixture \cite{tobe} also demonstrate
that solitons with large numbers of atoms require stronger attraction between
fermionic and bosonic components.

In conclusion, we have shown that bright solitons can be generated
in a Bose-Fermi mixture trapped in a three-dimensional elongated harmonic 
potential. For that the attraction between bosonic and fermionic atoms has 
to be strong enough changing in this way the repulsive interactions among
bosons to the attractive ones. Simultaneously, the attractive forces
induced in the bosonic cloud has to be weak enough to avoid the collapse.
It turns out that there exist windows in the values of the rubidium-potassium
scattering length within which the single-peak two-component structures neither 
spreading nor collapsing (therefore called by us solitons) can be formed. Those 
windows are shifting towards the region of less negative Bose-Fermi scattering 
length (weaker attraction) while elongating the trap making thus possible the 
formation of solitons even at the natural strength of the boson-fermion 
attraction.

\acknowledgments 
We are grateful to K. Bongs and M. Gajda for helpful discussions. We acknowledge 
support by the Polish Ministry of Scientific Research Grant Quantum Information and 
Quantum Engineering No. PBZ-MIN-008/P03/2003.

\end{document}